\documentclass[pre,aps,twocolumn, floatfix, showpacs]{revtex4}
\usepackage{amsmath,bm,graphicx}
\usepackage{amssymb}
\usepackage{amsfonts}
\newcommand{\V}[1]{{\bf #1}}

\newcommand{\myP}[3]{\phi_{#1\,#2\,#3}}
\newcommand{\pd}[2]{\frac{\partial #1}{\partial #2}}
\begin{document}

\title{Scaling properties of one-dimensional cluster-cluster aggregation with L\'{e}vy diffusion}

\author{Colm Connaughton}
\email{connaughtonc@gmail.com}
\author{Jamie Harris}

\email{Jamie.Harris@warwick.ac.uk}
\affiliation {Mathematics Institute and Centre for Complexity Science, University of Warwick, Gibbet Hill Road, Coventry CV4 7AL, UK}

\date{\today}

\begin{abstract}
We present a study of the scaling properties of cluster-cluster aggregation
with a source of monomers in the stationary state when the spatial transport
of particles occurs by L\'{e}vy flights. We show that the transition from
mean-field statistics to fluctuation-dominated statistics which, for the more
commonly considered case of diffusive transport, occurs as the spatial
dimension of the system is tuned through two from above, can be mimicked
even in one dimension by varying the characteristic exponent, $\beta$, of the 
the L\'{e}vy jump length distribution. We also show that the two-point
mass correlation function, responsible for the flux of mass in the stationary
state, is strongly universal: its scaling exponent is given by the mean field
value independent of the spatial dimension and independent of the
value of $\beta$. Finally we study numerically the two point spatial 
correlation function which characterises the structure of the depletion zone 
around heavy particles in the diffusion limited regime. We find that this
correlation function vanishes with a non-trivial fractional power of the
separation between particles as this separation goes to zero. We provide
a scaling argument for the value of this exponent which is in reasonable
agreement with the numerical measurements.
\bigskip

\noindent Citation: C Connaughton and J Harris J. Stat. Mech. (2010) P05003

\noindent doi: 10.1088/1742-5468/2010/05/P05003
\end{abstract}
\maketitle

\section{Introduction to cluster-cluster aggregation (CCA)}
\label{sec-intro}

Cluster-cluster aggregation (CCA) refers to the statistical dynamics of a
large collection of particles (or clusters) which move around under the action 
of some transport process and which, in addition, have a probability to 
coalesce irreversibly when 
they collide. The transport is usually stochastic, diffusion being a 
commonly studied case, and coalescence of clusters conserves mass. In general,
both the rate of coalescence and the transport coefficient are functions of
cluster mass.
Such dynamics are relevant to many physical phenomena. 
Commonly cited applications include the study of aerosols 
\cite{FRI2000}, the development of structure in the expanding universe 
\cite{SW1978}, clustering of algae \cite{ACK1997} and cloud droplet 
coalescence \cite{SCO1968}. From a theoretical perspective, CCA can be modelled
using relatively simple interacting particle systems and can thus provide a 
testing ground for statistical mechanics theories of complex systems.
The theoretical objective is to understand the statistical properties of
the mass distribution of the clusters.

CCA is an intrinsically non-equilibrium
process. Aggregation is irreversible so there is no possibility for
the statistical dynamics to have detailed balance. In some circumstances
it may be appropriate to consider fragmentation of clusters in addition to
aggregation. In such cases, detailed balance may be achievable. We do not 
consider such situations here however.
Two different scenarios may be studied which we refer to as the decay 
and forced cases. In the decay case, the more common of the two, the system is 
initially populated 
with a large number of small particles (monomers) which are considered to be the
fundamental unit for the aggregation process and whose mass can accordingly
be taken to be one. The size distribution subsequently spreads in the
space of cluster masses as larger clusters are generated by the aggregation of
smaller ones. This time evolution proceeds for  all time (we do not consider 
models exhibiting a gelation transition here) often in a self-similar fashion.
In the forced case, the system initially has no clusters but we continually
deposit monomers into the system at a constant rate from an external source. 
The size distribution again spreads in the mass space but, due to the
constant replenishment of the monomers by the external source, eventually
reaches a stationary state at any given cluster size. In this stationary state,
the depletion of clusters of a given size by aggregation events is exactly
balanced by the formation of such clusters by the aggregation of smaller
ones. This results in a constant flux of mass in the mass space. An analogy
may be drawn with the flux of energy through scales in the Kolmogorov
cascade of hydrodynamic turbulence \cite{CRZ2004}. We shall be concerned
primarily with the stationary forced case.

The most basic statistical measure is the average mass density,
$N(m,t)$, giving the average density of clusters of mass $m$ at time
$t$.  At the mean-field level,  $N(m,t)$ satisfies the Smoluchowski 
equation \cite{SMO1917}. For many applications, the collision kernel in
the Smoluchowski equation, which reflects the mass dependence of the 
underlying microscopic coalescence rates, is a homogeneous function whose
degree we shall denote by $\zeta$. For a broad class of such kernels,
the Smoluchowski equation has self-similar solutions which describe scaling
behaviour of $N(m,t)$ in both $m$ and $t$. In particular, $N(m,t)$ has power
law asymptotics for $N(m,t)$ as a function of mass. In the decay case,
we often find $N(m,t) \sim m^{-(\zeta+1)}$. In the forced case, the 
stationary state is given by $N(m) \sim m^{-(\zeta+3)/2}$ \cite{CS1998,CRZ2004}. 
The reviews by by Aldous \cite{ALD1999} and Leyvraz \cite{LEY2003} provide
comprehensive discussions of the Smoluchowski equation and its scaling
solutions.

One of the most interesting aspects of CCA is that this mean field description
can fail under certain conditions. This results in a transition to a fluctuation
dominated regime with completely different scaling properties for the large
$m$ asymptotics. The fundamental assumption underpinning mean field
theory is the absence of spatial correlations. In particular, the joint
distribution of clusters having masses $m_1$ and $m_2$ at a given point,
$\V{x}$, which is linked to the probability of collision, factorises as
the product of the one-point distributions of clusters with masses $m_1$ and
$m_2$. Likewise for higher order joint distributions. This works well in
the so-called reaction limited regime when the density of clusters is
sufficiently high that each aggregating cluster has a large number of potential
partners to with which to aggregate. In the most commonly
considered case, in which the transport process is simple diffusion, this
assumption turns out to be inconsistent for large masses if the spatial 
dimension is less than or equal to two. The reason is that in two dimensions 
and below
random walks are recurrent so that clusters which encounter each other once 
are highly probable to do so infinitely many times and hence to aggregate. 
Heavy clusters thereby generate an effective depletion zone around themselves.
The dynamics thus becomes diffusion limited and dominated by
diffusive fluctuations making the system much more difficult to analyse. The 
only case for which a
systematic study of the diffusion limited regime has been done is the
case of constant kernel in which the aggregation rate is 
independent of mass ($\zeta=0$). For the forced case in $d<2$, the diffusion 
limited stationary mass density can be shown to scale as
$N(m)\sim m^{-(2d+2)/(d+2)}$ which is shallower than the corresponding
mean-field prediction, $N(m)\sim m^{-3/2}$. Note that the diffusion limited
scaling exponent coincides with the reaction limited one when $d=2$ which
reflects the fact that $d=2$ is the critical dimension for this model. In 
addition to having a different value for the scaling exponent of
$N(m)$, the statistical properties of the diffusion limited regime are also
structurally different: higher order correlation functions exhibit 
multi-scaling in the sense that they do not scale as powers of the
corresponding single-point densities as they do in the mean field regime.

In this article we focus in detail on this transition from mean field to
fluctuation dominated regime in CCA. The fact that the parameter 
controlling the transition is the physical dimension of the model is 
inconvenient for both numerical exploration and physical applicability. We
remark that the important ingredient for the application of mean field
theory is not the physical dimension per se but rather whether or not
the transport mechanism has the ability to break spatial correlations 
between particles. We show that by replacing diffusive transport with
L\'{e}vy flights it is possible to continuously tune between 
fluctuation-dominated behaviour and mean-field behaviour {\em even in one
dimension}. The tuning parameter is the characteristic exponent of the 
L\'{e}vy distribution underlying the jump size distribution for particle
hops. The physical intuition is that correlations between particles are
broken even in low dimensions if there are sufficiently many long range
hops to effectively mix the clusters. We also explore the special role 
played in the forced case by the second order mass correlation function
whose scaling exponent is strongly universal as a result of the so-called
Constant Flux Relation (CFR) \cite{CRZ2007}. Its scaling exponent is, 
perhaps surprisingly, independent of both the spatial 
transport mechanism and the physical dimension and is always given by its mean
field value.

The outline of the article is as follows. We begin by defining the stochastic 
model used in studying CCA, as well as a brief overview of the statistical 
field theory used in deriving the Lee-Cardy equation for this model 
(Section 2). Then, we summarise the scaling properties of the multi-particle 
correlation functions or arbitrary order as applied to both reaction-limited 
and diffusion-limited regimes (Section 3). Using the Constant Flux Relation 
(CFR) we then derive the scaling invariance of the flux-carrying correlation 
function, demonstrating that this result is independent of the dimension of 
the system and of the mechanism of spatial transport (Section 4). Implementing 
L\'{e}vy flights for the one-dimensional model, we then characterise the 
suggested form of the mass density as a function of the L\'{e}vy exponent, 
which we validate against numerical simulation together with taking direct 
measurements for the integrated forms of the mass correlation functions 
(Section 5). Here we verify the scaling independence of the flux-carrying 
correlation function as we alter the spatial transport mechanism.  We finish 
the paper by determining a predicted form for the structure of the depletion 
zones as a function of the L\'{e}vy exponent (Section 6), which we validate 
numerically before providing a conclusion.

\section{The Lee-Cardy equation and stochastic Smoluchowski equation}
\label{sec-LeeCardy}

We now describe a simple stochastic model, first introduced without injection
of particles in \cite{KR1985} and with injection of particles in \cite{TNT1988},
which can be used to study the 
phenomenon of CCA. Discrete particles, each carrying a mass index, $m$,  occupy
 a $d$-dimensional square lattice, ${\mathbb L}$, with no restrictions on the number of 
particles occupying each lattice site. We take the mass index to be a
non-negative integer multiple of a fundamental mass, $m_0$. This means that all 
particles are composed of clusters of a single basic monomer unit. The state of the system at time $t$ is 
characterised by the set of occupation numbers, $N_{\V{x}\,m\,t}$, specifying
the number of particles of each mass $m$ at each point, $\V{x}$ of the lattice.
The occupation numbers evolve in time according to the following microscopic
dynamical transitions which occur in continuous time with exponentially 
distributed waiting times:

\begin{itemize}
\item Diffusion: A particle at any site may hop to an adjacent site. These
hops occur at rate $\widetilde{D}$, taken to be independent of particle mass.
\item Injection of monomers: A new monomer may appear at any site on ${\mathbb L}$.
The injection of monomers takes place at rate $\widetilde{J}$.
\item Aggregation: Two particles having masses $m_1$ and $m_2$ which are
at the same lattice point, $\V{x}$, may aggregate to create a single new particle having
mass $m_1+m_2$. This occurs at rate $\widetilde{\lambda} N_{\V{x}\,m_1\,t} ( N_{\V{x}\,m_2\,t} - \delta_{m_1\,m_2})$ where $\widetilde{\lambda}$ is the
reaction rate, taken to be independent of mass, and the $N_{\V{x}\,m\,t}$ 
factors encode the pairwise nature
of the aggregation process (the Kronecker symbol, $\delta_{m_1\,m_2}$, ensures
that the rate is correct when two particles of the same mass aggregate).
\end{itemize}
The ``rates" in these rules refer to the corresponding waiting times. These
rules define a Markov chain which can be easily simulated using the Gillespie
algorithm  \cite{GIL1977}. All numerical results presented in this paper have 
been obtained in this fashion.

This model has a Master Equation whose solution
can be used to compute averages with respect to the microscopic dynamics. Such
averages, which we shall denote by $\langle . \rangle_{\mathbb L}$, are
 equivalent to averages with respect to the effective action of an associated
statistical field theory. This effective action is obtained from the
dynamical rules by the well known Doi-Peliti construction. For details of the
procedure for this specific model see \cite{CRZ2006}. For many microscopic
models describing interacting particle systems, averages with respect to this effective action are further equivalent to averages with respect
to solutions of a stochastic rate equation \cite{LEE1994,CAR1998} known
as the Lee-Cardy equation. The Lee-Cardy equation specific to our CCA model, 
first derived in \cite{ZAB2001}, takes the form of a stochastic Smoluchowski
equation. In the continuous limit it takes the form:

\begin{eqnarray}
\nonumber
\partial_t\myP{\V{x}}{t}{m} &=& D\Delta \myP{\V{x}}{t}{m} + \frac{J}{m_0}\delta(m-m_0)\\
\nonumber
&+& \lambda\int_{0}^{m}\myP{\V{x}}{m'}{t}\, \myP{\V{x}}{m-m'}{t}\,dm' \\
\nonumber
&-& 2\lambda\,\myP{\V{x}}{m}{t}\,\int_{0}^{\infty}\myP{\V{x}}{m'}{t}\,dm'\\
\label{eq-SSE}
&+&i\,\sqrt{\lambda}\,\myP{\V{x}}{m}{t}\,\xi_{\V{x}\,t}
\end{eqnarray}

where $\xi_{\V{x}\,t}$ is Gaussian white noise and $\Delta$ is the 
$d$-dimensional
Laplacian. $D$, $\lambda$ and $J$ are respectively the diffusion constant, 
reaction rate and mass injection rate obtained from an appropriate rescaling
of the corresponding microscopic rates with the lattice spacing in the
continuous limit \cite{CRZ2006}. We shall denote averages with respect to 
the solutions of Eq.~(\ref{eq-SSE}) by  $\langle . \rangle_\xi$.

One of the most striking aspects of the Lee-Cardy equation is that the noise
term is imaginary indicating that the field $\myP{\V{x}}{m}{t}$  cannot
be literally interpreted as the density of particles, $N_{\V{x}\,m\,t}$, 
however tempting such an identification may seem. Imaginary noise encodes the 
fact that the
particles in the system are anti-correlated \cite{LEE1994,HT1997}. Physically,
this anti-correlation comes about since the process of aggregation tends, on 
average, to produce a depletion zone around each particle. We shall discuss
these depletion zones in detail in Sec.~\ref{sec-anticorrelations}.

Careful study of the construction leading to Eq.~(\ref{eq-SSE}) leads 
to the conclusion \cite{CRZ2006} that single point {\em moments} of $\myP{\V{x}}{m}{t}$ 
correspond to {\em factorial moments} of the physical density, 
$N_{\V{x}\,m\,t}$:
\begin{equation}
\label{eq-PhiNRelation}
\langle \myP{\V{x}}{m}{t}^n \rangle_\xi = \frac{1}{n!} \langle N_{\V{x}\,m\,t}(N_{\V{x}\,m\,t} -1 )\ldots(N_{\V{x}\,m\,t} -n +1)\rangle_{\mathbb L}.
\end{equation}
Multi-point moments translate directly. For example,
\begin{equation}
\label{eq-PhiNRelation2}
\langle \myP{\V{x}}{m_1}{t} \myP{\V{x}}{m_2}{t} \rangle_\xi = \langle N_{\V{x}\,m_1\,t}N_{\V{x}\,m_2\,t}\rangle_{\mathbb L} \mbox{ if $m_1 \neq m_2$.}
\end{equation}
These relations allow us to translate correlation functions of 
Eq.~(\ref{eq-SSE}) which are most suitable for theoretical analysis into 
corresponding correlation functions in the original lattice model which are
most suitable for numerical measurements. In relation to the interpretation
of the field $\myP{\V{x}}{m}{t}$ alluded to above, Eq.~(\ref{eq-PhiNRelation})
tells us that $\myP{\V{x}}{m}{t}$ is equal to the density, $N_{\V{x}\,m\,t}$,
on average but differs in distribution.

If the occupation numbers are large, we can, to leading order,  ignore the 
distinction between moments and factorial moments on the right hand side of 
Eq.~(\ref{eq-PhiNRelation}). Furthermore, if we could neglect correlations 
between particles, higher order correlation functions could be factorised 
into products of densities:
\begin{equation}
\label{eq-meanFieldApprox}
\langle N_{\V{x}\,m_1\,t}N_{\V{x}\,m_2\,t}\rangle_{\mathbb L} \approx \langle N_{\V{x}\,m_1\,t}\rangle_{\mathbb L} \langle N_{\V{x}\,m_2\,t}\rangle_{\mathbb L}.
\end{equation}
This is the essence of the mean field approximation for CCA.  If the density
exhibits scaling with some physical quantity such as the mass or the
source strength, it follows from the mean field approximation that the 
higher order correlation functions exhibit simple scaling. That is to say
that if the density has a scaling exponent, $\gamma_1$, then the corresponding
scaling exponent of an order $n$ correlation function is $\gamma_n=n\,\gamma_1$.

We have already mentioned that the interactions between particles resulting 
from aggregation provides a natural mechanism for generating anti-correlations
between particles. The extent to which the approximation 
(\ref{eq-meanFieldApprox}) is a good one therefore depends on the strength of
these anti-correlations. It has been understood that in low dimensions  the large
time, large mass behaviour is always diffusion limited resulting in
anticorrelations between particles which are sufficiently strong to 
invalidate the mean field approximation. Low dimensions in this case means $d<2$ with
the case $d=2$ being marginal.  One expects to observe a violation of mean 
field theory accompanied by the onset of multiscaling as one varies
the dimension of the lattice ${\mathbb L}$. This is
indeed true with strong multiscaling observed in both the decaying  
\cite{ZAB2001} and forced \cite{CRZ2005,CRZ2006} cases in $d=1$ which mellows to
logarithmic corrections to the mean-field scaling in $d=2$.   
Triggering this transition from mean-field to non-mean-field behaviour by
varying the spatial dimension is not very easy in practice. 
In Sec. \ref{sec-Levy} of this article we 
show that introducing long-range diffusion can break the anti-correlations 
between particles resulting in behaviour which is very reminiscent of
mean-field behaviour {\em even in one dimension}. Before this however we
will review the scaling properties of CCA with regular diffusion.

\section{Scaling properties of CCA with a source of monomers: reaction limited and diffusion limited regimes}
\label{sec-CCAWithSource}

A constant injection of monomers into the system allows for the creation of a 
non-equilibrium stationary state, which is characterised by a constant flux of 
mass from smaller to larger masses under aggregation. However, since there is no upper 
mass limit defined on the system, the creation of arbitrarily large clusters
can go on indefinitely. In this sense, the stationary state is only defined 
for some inertial range of masses. Any stationary solution to the Stochastic 
Smoluchowski Equation describing the mass density must therefore be of 
Kolmogorov type \cite{CRZ2004}. That is, these solutions exhibit scaling 
evocative of the Kolmogorov spectrum found in forced hydrodynamic turbulence, 
with a cascade defined by a constant flux of mass from smaller to larger 
clusters mediated by the aggregation of particles. From this point on, we 
shall assume statistical homogeneity and drop the $\V{x}$ index.

The mass density has dimension $[N_m] = L^{-d}\, M^{-1}$. The 
dimensions of the physical parameters appearing in Eq.~(\ref{eq-SSE})
are as follows $[D]=L^2\,T^{-1}$, $[\lambda]=L^d\,T^{-1}$ and 
$[J] = M\,L^{-d}\,T^{-1}$. In order to describe a cascade, the stationary
density should depend on the mass flux, $J$. There are two natural ways to
construct an expression with the correct dimension using only one additional
parameter in addition to the flux:
\begin{eqnarray}
\label{eq-MFDensity} N_m &\propto& \sqrt{\frac{J}{\lambda}}\,m^{-\frac{3}{2}},\\
\label{eq-DLDensity} N_m &\propto& \left(\frac{J}{D}\right)^{\frac{d}{d+2}}\,m^{-\frac{2d+2}{d+2}}.
\end{eqnarray}
The first corresponds to the reaction-limited regime in which diffusion
plays no role. The second 
corresponds to the diffusion-limited regime in which the reaction rate
plays no role since reactions are effectively instantaneous compared to
the time required for particles to find each other by diffusion. 

If we average Eq.~(\ref{eq-SSE}) in the statistically homogeneous case and
make the mean field approximation, Eq.~(\ref{eq-meanFieldApprox}), then
we obtain the standard Smoluchowski coagulation equation:
\begin{eqnarray}
\nonumber \partial_t N_m(t) &=& \lambda \int_0^{\infty}dm_1 dm_2 N_{m_1} N_{m_2} \delta(m-m_1-m_2)\\
\nonumber &-& \lambda \int_0^{\infty}dm_1dm_2 N_{m} N_{m_1}\delta(m_2-m-m_1)\\
\nonumber &-& \lambda \int_0^{\infty}dm_1dm_2 N_{m} N_{m_2}\delta(m_1-m_2-m)\\
\label{eq-Smoluchowski}&+& (J/m_0)\ \delta(m-m_0).
\end{eqnarray}
The exact stationary solution of this equation is \cite{CS1998,CRZ2004} 
\begin{equation}
\label{eq-exactMFDensity}
N_m = \sqrt{\frac{J}{2\pi\lambda}}\,m^{-\frac{3}{2}},
\end{equation}
confirming that the reaction-limited scaling, Eq.~(\ref{eq-MFDensity}), 
corresponds to the mean field scaling. In one dimension, exact calculations
\cite{TNT1988,HUB91} showed that for the case of instantaneous reactions, 
$\lambda\to \infty$, 
\begin{equation}
\label{eq-exactDLDensity}
N_m \sim \Gamma(1/3)^{-1}\left(\frac{4 J}{9 D}\right)^\frac{1}{3}\, m^{-\frac{4}{3}},
\end{equation}
for $m \gg 1$, which agrees with the diffusion-limited scaling, Eq.~(\ref{eq-DLDensity}), 
with $d=1$. Furthermore the dynamical scaling properties of the 
time-relaxation to these 
steady states in both the mean-field and one-dimensional cases were 
computed exactly in \cite{MS1993}.
It was subsequently shown in  \cite{RM2000} that
the diffusion-limited scaling, Eq.~(\ref{eq-DLDensity}), is actually 
exact for all $d<2$.
The fact that the exponents in Eqs.~(\ref{eq-MFDensity}) and 
(\ref{eq-DLDensity}) become equal when $d=2$ reflects that the critical 
dimension for this system is 2. 

The presence of multiscaling in $d<2$ can be detected 
by measuring the higher order mass correlation functions:
\begin{equation}
c_n(m_1,\ldots,m_n, t) = \langle \myP{\V{x}}{m_1}{t} \ldots  \myP{\V{x}}{m_n}{t} \rangle_\xi.
\end{equation}
The case $n=1$ is just the mass density as discussed above. In the 
statistically homogeneous case, these correlation functions do not
depend on the spatial point, $\V{x}$, at which they are measured. For 
aggregation with source, $c_n(m_1,\ldots,m_n, t)$ becomes independent of $t$ 
as $t\to \infty$ and is a homogeneous function of mass for large masses:
\begin{equation}
c_n(h\,m_1,\ldots,h\,m_n) \sim h^{\gamma_n} c_n(m_1,\ldots,m_n).
\end{equation}
In \cite{CRZ2005}, the exponents $\gamma_n$ were calculated perturbatively
in $\epsilon=2-d$ using dynamical renormalisation group:
\begin{equation}
\label{eq-gamma_n}
\gamma_n = \left(\frac{2d+2}{d+2}\right)n + \frac{n\,(n-1)}{2\,(d+2)}\epsilon + O(\epsilon^2),
\end{equation}
and found to agree remarkably well with Monte Carlo simulations in $d=1$ 
despite the fact that, formally, $\epsilon=1$ in this case.  The epsilon
expansion written in this way, with $d$ appearing explicitly,
demonstrates that the effect of fluctuations is to correct the
{\em simple} scaling $\gamma_n = n\,\gamma_1$ (recall that $\gamma_1$ is 
given exactly by $\gamma_1 = \frac{2 d + 2}{d+2}$ \cite{RM2000}), thereby
demonstrating the presence of multi-scaling in the model. Full details are
given in \cite{CRZ2006}. Direct measurement
of the exponents $\gamma_n$ from numerics is quite difficult for $n>1$ due to
the strong fluctuations in the diffusion-limited regime. In practice, 
for $n>1$, better statistics are obtained by measuring the integrated 
correlation functions,
\begin{equation}
\label{eq-integratedCorrelationFunction}
C_n(m) = \int_m^\infty\!\!\! dm_1 \ldots  \int_m^\infty\!\!\! dm_{n-1}\, c_n(m, m_1\ldots,m_{n-1}).
\end{equation}
$C_n(m)$ can be expressed in terms of site occupation numbers using 
Eqs.~(\ref{eq-PhiNRelation}) and (\ref{eq-PhiNRelation2}) and scales with 
exponent $\gamma_n-n+1$. Notice from Eq.~(\ref{eq-gamma_n}) that
$\gamma_2=3$, independent of the spatial dimension, to first order in the 
$\epsilon$-expansion. Furthermore, this is exactly the exponent one would
expect from  mean-field theory even though Eq.~(\ref{eq-gamma_n}) 
specifically describes the non-mean-field regime. In fact, $\gamma_2$ is
exactly given by its mean-field value in all dimensions regardless of
whether mean-field theory applies or not. This is a specific case of
a general property of many non-equilibrium stationary states which carry a
flux of some conserved quantity: strong universality of the flux-carrying 
correlation function \cite{CRZ2007}. The fact that conservation laws 
determine the scaling of the flux-carry correlation function exactly
is well know in turbulence as Kolmogorov's $\frac{4}{5}$-Law (see \cite{FRI1995} for a review and wide-ranging discussions). We now study this important 
feature in more detail.

\section{Constant Flux Relation}
\label{sec-CFR}
\begin{figure}
\centering
\includegraphics[width=7cm]{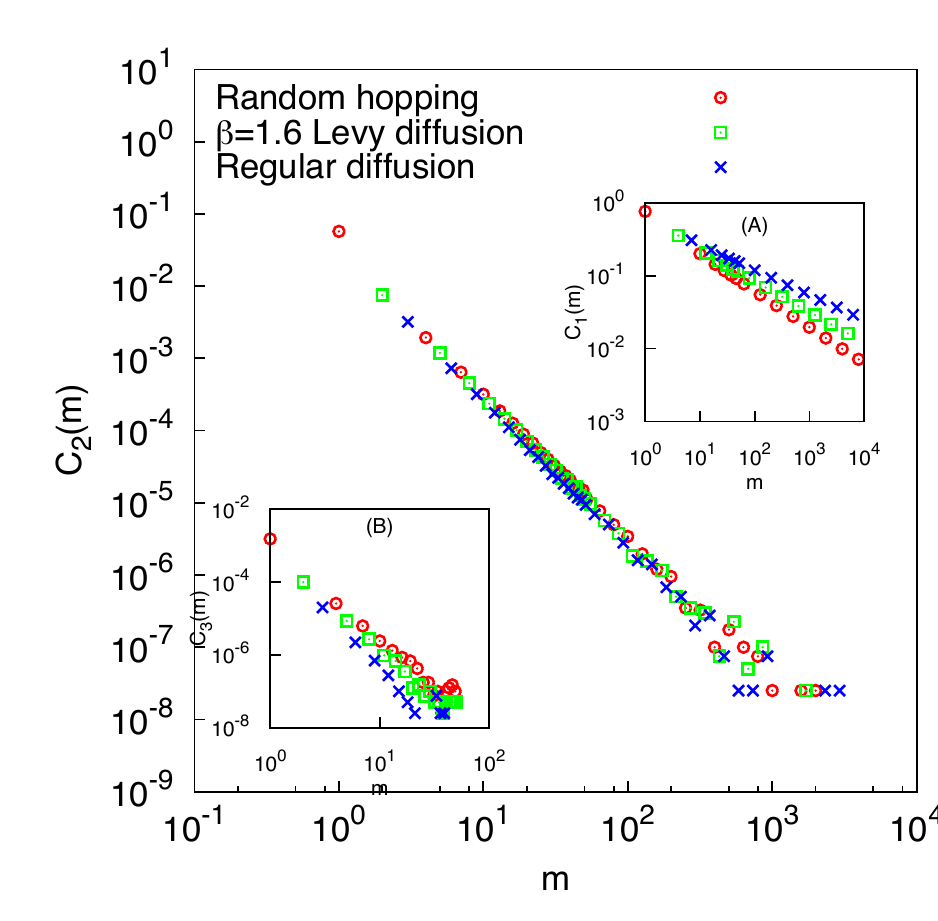}
\caption{\label{fig-scaling} Scaling universality of the second order integrated mass correlation function, $C_2(m)$ as given by Eq. (\ref{eq-integratedCorrelationFunction}), with mass, $m$, as shown for three different transport mechanisms. In comparison, scaling for the first and third order integrated mass correlation functions are shown in insets (A) and (B) respectively.  }
\end{figure}

We have already remarked that in the forced case, CCA reaches a 
quasi-stationary state characterised by a constant flux of mass from small to 
large masses.  The Constant Flux Relation (CFR) \cite{CRZ2007} expresses the
fact that this constant flux of mass exactly determines the scaling of the 
flux-carrying correlation function in the stationary state, a common feature
of most driven non-equilibrium systems which reach stationary states dominated
by a constant flux of a conserved quantity. 

The Lee-Cardy equation, Eq.~(\ref{eq-SSE}), is equivalent to the underlying 
effective field theory, a fact which sometimes calls into question the
usefulness of the Lee-Cardy formalism. The derivation of the CFR, however,
 is enormously easier using the Lee-Cardy equation than it would be otherwise.
At stationarity, and assuming spatial homogeneity, Eq.~(\ref{eq-SSE}) can be 
written
\begin{eqnarray}
\label{unsym}
\nonumber
0&=&\lambda\int_{0}^{m}\langle\phi(m')\phi(m-m')\rangle_\xi dm'\\
&-&2\lambda\int_{0}^{\infty}\langle \phi(m)\phi(m')\rangle_\xi dm',
\end{eqnarray}
where we have taken averages with respect to the noise term $\xi$ and let $m>m_0$. Let us assume that for large masses, the correlation function $\langle \phi(m_1)\phi(m_2)\rangle_\xi$ takes the form
\begin{equation}
C(m_1,m_2) = (m_1m_2)^{-x}\psi\left(\frac{m_1}{m_2}\right),
\end{equation}
where we have introduced the dimensionless scaling function $\psi$.
By the symmetry of $C$, $\psi$ has the property $\psi(x)=\psi(x^{-1})$. This 
defines our flux-carrying correlation function. Re-writing Eq.~(\ref{unsym}) 
in the symmetric form
\begin{eqnarray}
\label{symmetric}
\nonumber
0 &=& \lambda\int_{0}^{\infty}\int_{0}^{\infty}C(m_1,m_2)\delta(m-m_1-m_2)dm_1dm_2\\
\nonumber
&-&\lambda\int_{0}^{\infty}\int_{0}^{\infty}C(m,m_1)\delta(m_2-m-m_1)dm_1dm_2\\
\nonumber
&-&\lambda\int_{0}^{\infty}\int_{0}^{\infty}C(m,m_2)\delta(m_1-m_2-m)dm_1dm_2,\\
&&
\end{eqnarray}
and applying the Zakharov transformations \cite{ZLF92}
\begin{eqnarray}
\nonumber
\nonumber(m_1,m_2)&\longrightarrow\left(\frac{mm_1}{m_2},\frac{m^2}{m_2}\right),\\
\label{Ztrans}
(m_1,m_2)&\longrightarrow\left(\frac{m^2}{m_1},\frac{mm_2}{m_1}\right)
\end{eqnarray}
to the second and third integrals of Eq.~(\ref{symmetric}) respectively, 
results in an expression in which each delta-function has the same argument. 
This gives the condition
\begin{eqnarray}
\nonumber
0&=&\int_{0}^{\infty}\int_{0}^{\infty}(m_1m_2)^{-x}\psi\left(\frac{m_1}{m_2}\right)\left(m^{y}-m_2^{y}-m_1^{y}\right)\\
&&\qquad\qquad\qquad\times\;\delta(m-m_2-m_1)dm_1dm_2,
\end{eqnarray}
which can only be satisfied for $y:=2x-2 = 1$, or equivalently, 
$x=\frac{3}{2}$. Therefore, the flux-carrying correlation function must be of 
the form
\begin{equation}
\label{corr}
C(m_1,m_2) \sim (m_1m_2)^{-\frac{3}{2}}\psi\left(\frac{m_1}{m_2}\right)
\end{equation}
for sufficiently large masses, which is a homogeneous function of degree -3.
We have implicitly assumed that the corresponding integrals in Eq.~(\ref{symmetric})
converge for this exponent. This can be demonstrated explicitly for the 
constant kernel case but, for general kernels must be verified numerically
\cite{CRZ2008}.
An immediate corollary of this result applies to the specific case $m_1 = m_2$: 
using the factorial moments relationship in Eq. (\ref{eq-PhiNRelation}) we 
have that
\begin{equation}
\label{scaleinv}
\frac{1}{2}\langle N_m(N_m-1)\rangle_{\mathbb{L}} \sim m^{-3}.
\end{equation}
Namely, that the stationary average density of pairs must scale with exponent 
-3 for sufficiently large $m$.

In the mean field regime, higher order correlation functions
factorise into products of densities as expressed in 
Eq.~(\ref{eq-meanFieldApprox}). It immediately follows from 
Eq.~(\ref{eq-exactMFDensity}) that $\gamma_2 = 3$ in mean field theory. 
Thus the scaling exponent, $\gamma_2$, is given by its mean field 
value in all dimensions. We note also that the dependence on the transport 
process disappears on averaging Eq.~(\ref{eq-SSE}) so that the exponent
$\gamma_2$ is strongly universal, as suggested by figure (\ref{fig-scaling}). In the next section we consider CCA with
L\'{e}vy flights. The choice of the characteristic exponent, $\beta$, of the
L\'{e}vy flights strongly affects the scaling of general correlation
functions in the model. We shall see, however, that the value of $\gamma_2$
is insensitive to the choice of $\beta$.

\section{One-dimensional CCA with L\'{e}vy Diffusion}
\label{sec-Levy}

\begin{figure}
\centering
\includegraphics[width=7cm]{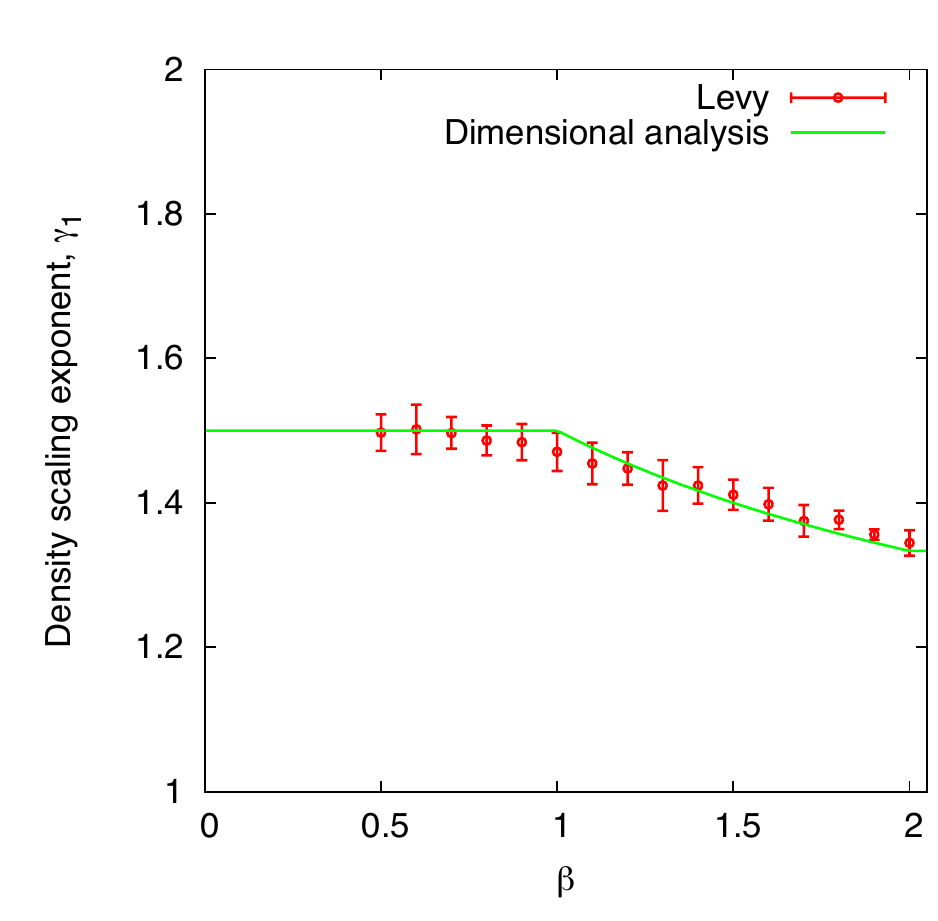}
\caption{Comparison of the numerically measured 
exponent $\gamma_1$ against the prediction derived from dimensional analysis, Eq.~(\ref{eq-gamma1Levy}), as plotted against the L\'{e}vy characteristic, $\beta$ . Data was taken from stationary simulations on a lattice of size $10^5$. The error bars 
on the exponents represent two standard deviations as estimated by 
bootstrapping a least squares estimator with the data measuring the integrated 
density, $C_1(m)$, as defined by Eq.~(\ref{eq-integratedCorrelationFunction}),
over 4 decades of $m$.}
\label{fig-dimensionalArgument} 
\end{figure}

\begin{figure}
\centering
\includegraphics[width=6.78cm]{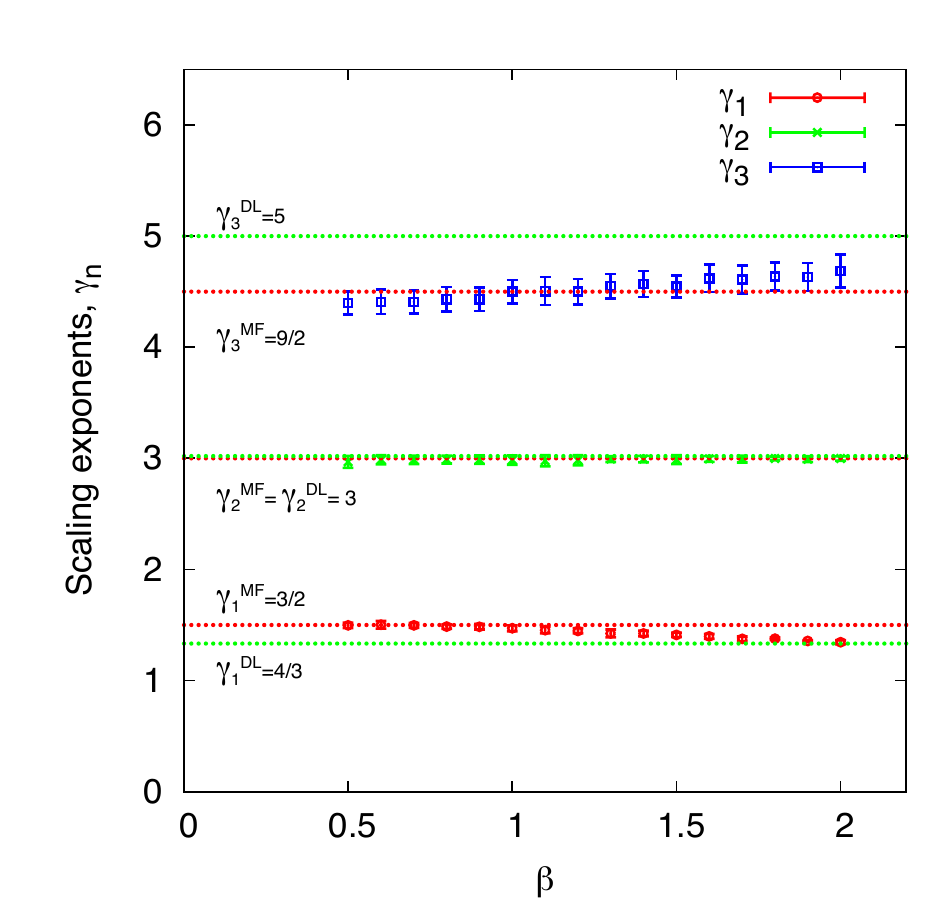}
\caption{Scaling exponents for the first three mass
correlation functions as functions of L\'{e}vy characteristic $\beta$. The exponents $\gamma_2$
and $\gamma_3$ were estimated using a maximum likelihood estimator 
\cite{CSN2009} for a power law distribution of pairs and triplets entering
Eq.~(\ref{eq-integratedCorrelationFunction}). Theoretically predicted values for both the mean field and diffusion limited cases are included for each exponent. Error bars correspond to two standard deviations calculated by bootstrapping the
estimator.}
\label{fig-exponents} 
\end{figure}

CCA with L\'{e}vy diffusion is of some interest in certain surface growth phenomena \cite{AFH1998,LGV2008}
but our interest stems from the theoretical point of view that it allows
us to mimic the transition to mean-field behaviour even in one dimension. From
now on we focus entirely on the one dimensional case.

We modify the model introduced earlier by allowing particles to make long
range jumps. The waiting times between particle hops 
remain exponentially
distributed with a fixed rate. Rather than simply hopping to nearest
neighbours, however, the length of each jump is independently sampled from a 
probability distribution with  a heavy tail. In our numerical experiments,
we took this distribution to be a symmetric L\'{e}vy distribution with
scale factor 1 and scaling exponent $\beta\in (0,2]$. Jump lengths were 
rounded to the nearest integer to remain consistent with our lattice 
formulation. Such hops are referred to as L\'{e}vy {\em flights} \cite{MK2000}
and should not be confused with the (often more physically realistic)
case of L\'{e}vy {\em walks} in which the waiting time distribution and the
jump length distribution are not independent to account for the fact that
for particles of finite mass long jumps should take more time. The probability
distribution, $P(x,t)$ of the position of a particle starting at $x=0$ at $t=0$
and exhibiting one dimensional L\'{e}vy flights with exponent $\beta$ satisfies
 a fractional diffusion equation: 
\begin{equation}
\label{eq-fractionalDiffusion}
\pd{P}{t} = D_\beta \frac{\partial^\beta P}{\partial x^\beta}
\end{equation}
where the generalised (or anomalous) diffusion coefficient, $D_\beta$, has dimension 
$L^\beta T^{-1}$ and the fractional derivative is defined through its Fourier
space representation. See  \cite{MK2000} for details. For $\beta=2$ the 
L\'{e}vy distribution reduces to the Gaussian distribution and we recover
the case of regular diffusion. In this case, the mean square 
displacement of a particle grows linearly in time $\langle x^2(t)\rangle = D t$
providing a natural way to define the diffusion length, 
$l_D = \sqrt{D\,t}$.  For $\beta<2$, $\langle x^2(t)\rangle$ is infinite for 
any finite time and we require a different approach to defining the anomalous
diffusion length. One reasonable way to do this uses the fact that
Eq.~(\ref{eq-fractionalDiffusion}) has a self-similar solution describing the
evolution of $P(x,t)$ in which the self-similar variable is $z=\frac{\left| x\right|}{(D_\beta\,t)^{1/\beta}}$ \cite{MK2000}. It therefore makes sense,
as one might expect from dimensional considerations, to define the anomalous 
diffusion length as $l_\beta = (D_\beta\,t)^{1/\beta}$ despite the fact that
the mean square displacement is divergent.

One may repeat the dimensional argument outlined in Sec.~\ref{sec-CCAWithSource}
in the diffusion-limited regime taking the anomalous diffusion coefficient, 
$D_\beta$, instead of $D$ and arrive for $d=1$ at 
\begin{equation}
\label{eq-DALevy}
N_m \propto \left(\frac{J}{D_\beta}\right)^{\frac{1}{1+\beta}}\,m^{-\frac{2+\beta}{1+\beta}}.
\end{equation}
For $\beta=2$ we recover the known diffusion limited scaling,
Eq.~(\ref{eq-exactDLDensity}). For $\beta=1$, the scaling becomes that of the
mean-field answer, Eq.~(\ref{eq-exactMFDensity}), with $D_\beta$ replacing
the reaction rate, $\lambda$. This suggests that at $\beta=1$, the L\'{e}vy
flights have become sufficiently long range to break all local
anti-correlations between particles. If $\beta$ is decreased below 1,
we do not expect that the scaling properties would change any further since
particles are now effectively independent. Thus we have the following
dimensional prediction for the scaling of the mass density in the
presence of L\'{e}vy flights:
\begin{equation}
\label{eq-gamma1Levy}
\gamma_1 = \left\{
\begin{array}{ll}
\frac{3}{2} & \beta <1\\
\frac{2+\beta}{1+\beta}&1\leq\beta\leq 2
\end{array}
\right.
\end{equation}

Fig.~\ref{fig-dimensionalArgument} compares the results of numerical 
measurements of $\gamma_1$ with the predictions of Eq.~(\ref{eq-gamma1Levy})
and indicates fair agreement. 
If we are correct in interpreting  Eq.~(\ref{eq-gamma1Levy}) as describing
the breaking of correlations by the long-range hops then it should follow 
that multiscaling described by Eq.~(\ref{eq-gamma_n}) should also be
lost as $\beta$ is tuned from $2$ to 1 and simple scaling should be
restored for the higher order correlation functions. This is illustrated
in Fig.~\ref{fig-exponents} which shows the measured scaling exponents of
the first three higher order correlation functions as a function of $\beta$.
We see that within error, simple scaling is restored in one dimension 
as $\beta\to 1$. This is not because we force the system to become
reaction-limited by increasing the number of particles per site
(the average number of particle per site in all simulations shown
here was about 0.8 so we remain in the low density regime).
Rather the transport process
breaks correlations between particles allowing the mean field behaviour 
to become observable again. It was pointed out to us that a somewhat similar
transition in the scaling exponents of CCA as a additional 
scaling parameter is varied was demonstrated within mean field theory in
\cite{CS1998}. There, an additional scaling parameter was introduced by 
the inclusion of exogeneous growth of clusters on top of the basic
model dynamics.

As an aside, we remark that the dimensional argument
leading to Eq.~(\ref{eq-gamma1Levy}) suggests
$\gamma_1 = \frac{2d+\beta}{d+\beta}$ when $d$ is left arbitrary. Setting 
$\gamma_1=\frac{3}{2}$ might lead one to conjecture that the critical 
dimension for CCA with L\'{e}vy diffusion is $d_c=\beta$. While intriguing,
this suggestion of a fractional critical dimension lacks the exact expression
for $\gamma_1$ which allowed a convincing argument to be made for $d_c=2$
in the case of regular diffusion ($\beta=2$) \cite{RM2000}.

\section{Structure of the depletion zone}
\label{sec-anticorrelations}

\begin{figure}
\centering
\includegraphics[width=7cm]{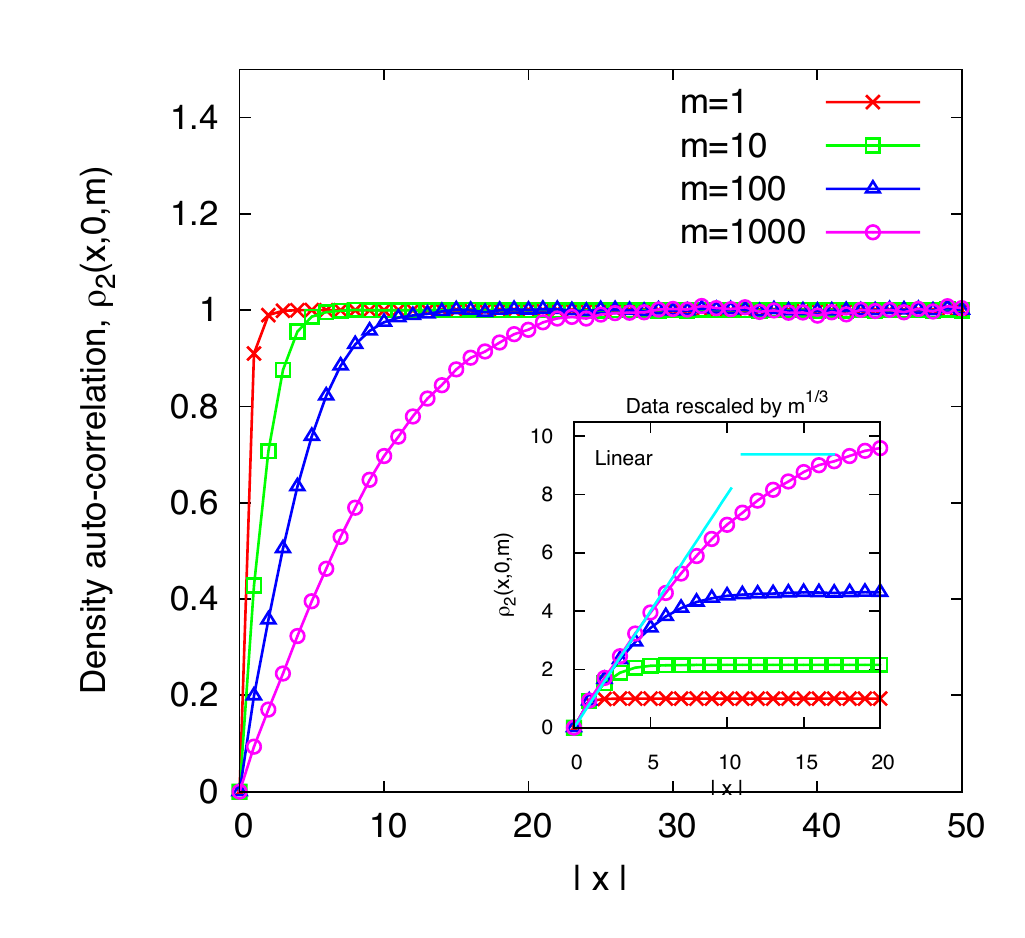}
\caption{The structure of the depletion zone as given by Eq. (\ref{eq-rhonCCA}) in the case of regular diffusion. These plots have been normalised against the single-point densities corresponding to each choice of mass, $m$. The inset shows plausible evidence in support of the predicted form given by Eq. (\ref{eq-rhonCCA}) after an appropriate rescaling of the data by $m^{1/3}$. }
\label{fig-anticorrelationVaryM}
\end{figure}

The breaking of anti-correlations between particles by L\'{e}vy flights should
also be seen in the structure of the depletion zones surrounding particles. 
We now turn our attention to characterising this effect in one dimension.
To detect the presence of depletion zones around particles in CCA we need to
calculate the multipoint correlation functions between densities at different spatial
points:
\begin{equation}
\rho_n(\V{x}_1,\ldots\V{x}_n,t) = \langle \myP{\V{x}_1}{m_1}{t} \ldots  \myP{\V{x}_n}{m_n}{t} \rangle_\xi.
\end{equation}
A closely related 2-point correlation function was calculated exactly
in one dimension for the case of infinite reaction rate in \cite{RM2000}
and demonstrated clearly the tendency for clusters to anti-correlate in
the diffusion-limited regime.
It is natural to normalise such correlation functions with the mass
density so that the ratio goes to 1 when particles are uncorrelated.
To avoid excessively clumsy notation, we suppress the mass dependence of the
spatial correlations for the time being. 

If we forget about the mass dependence of the particles for the moment 
we are left with the coagulation process $A+A\to A$. One can study how the
analogous correlation functions in this system behave. For the $A+A\to A$ 
system without source it was predicted in 
\cite{MRZ2006} and later proven rigorously in \cite{MRTZ2006} that in one
dimension:
\begin{equation}
\label{eq-AA2A}
\rho_n(\V{x}_1,\ldots\V{x}_n,t) \sim \left(\frac{1}{\sqrt{D\,t}}\right)^n \prod_{1\leq i < j \leq n} \frac{\left|\V{x}_i-\V{x}_j\right|}{\sqrt{D\,t}},
\end{equation}
where $\sim$ denotes the short-distance asymptotic behaviour.  The first term on 
the right hand side is simply the $n$th power of the density.  The second term
accounts for correlations, producing multiscaling of the time-decay
exponents for the $A+A\to A$ problem in $d=1$. The fact that  
$\rho_n(\V{x}_1,\ldots\V{x}_n,t)$ vanishes as any two points are  brought 
together reflects the fact that particles are anti-correlated.

Since the mechanism for the generation of anticorrelations in the stationary
CCA model is the same as for the $A+A\to A$, one can use Eq.~(\ref{eq-AA2A})
to guess the corresponding formula for the CCA model in the stationary state  \cite{CRZ2009}. In Eq.~(\ref{eq-AA2A}) the diffusion length for decaying
coagulation appears as $l_D = \sqrt{D\,t}$. The corresponding diffusion length 
for particles of mass $m$ in {\em stationary} CCA is, from dimensional 
considerations, 
\begin{equation}
l_D = \left(\frac{D\,m}{J}\right)^{\frac{1}{3}}.
\end{equation}
Upon substitution into Eq.~(\ref{eq-AA2A}) we arrive
at the following suggestion for the asymptotic behaviour, now in mass,  of
the multi-point correlation functions in stationary CCA:
\begin{equation}
\label{eq-rhonCCA}
\rho_n(\V{x}_1,\ldots\V{x}_n,m) \sim \left(\frac{J}{D}\right)^\frac{n}{3}\,m^{-\frac{4}{3}n} \prod_{1\leq i < j \leq n} \frac{\left|\V{x}_i-\V{x}_j\right|}{(\frac{D\,m}{J})^{1/3}}.
\end{equation}
In the absence of a complete calculation, the plausibility of this argument
should be assessed by comparison with numerics. This is done for $n=2$ in
Fig.~\ref{fig-anticorrelationVaryM}. Eq.~(\ref{eq-rhonCCA}) implies that
after rescaling with $m^{1/3}$, $\rho_2(\V{x}_1,\V{x}_2,m)/ \rho_1(m)^2$
for different values of $m$ should collapse onto the same curve which vanishes
linearly as the separation, $\left|\V{x}_1-\V{x}_2 \right|$, tends to $0$. 
Fig. ~\ref{fig-anticorrelationVaryM} clearly shows that the depletion
zone is larger for the heavier particles. The inset provides plausible 
evidence in support of the expected scaling.

\begin{figure}
\centering

\label{fig-anticorrelationVaryBeta}
\includegraphics[width=7cm]{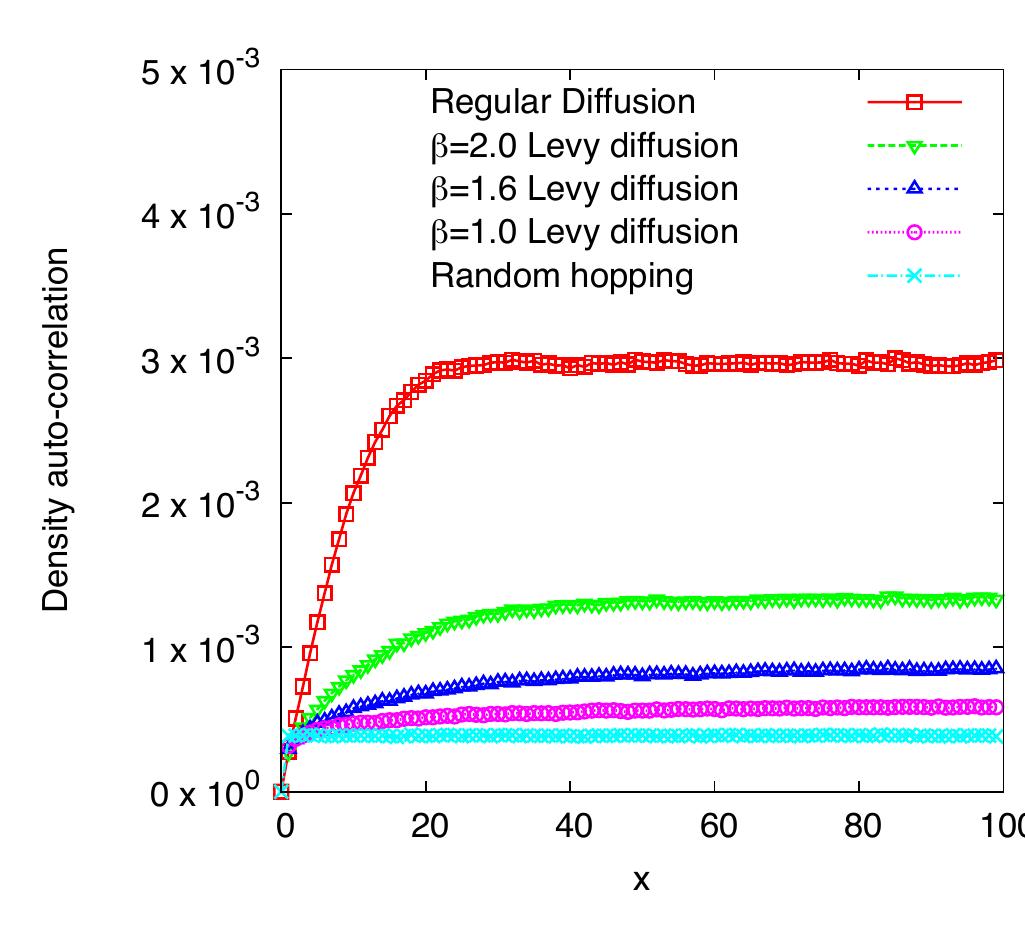}
\caption{The depletion zone as shown for different
transport mechanisms without normalisation against the single-point densities, for a fixed choice of mass, $m=1000$. As the L\'{e}vy characteristic, $\beta$ is varied from 2 down to 1, the anticorrelation void disappears, with the limiting cases of random lattice hops and regular diffusion also being shown.}

\end{figure}

\begin{figure}
\centering
\label{fig-newCorrelationHole}
\includegraphics[width=7cm]{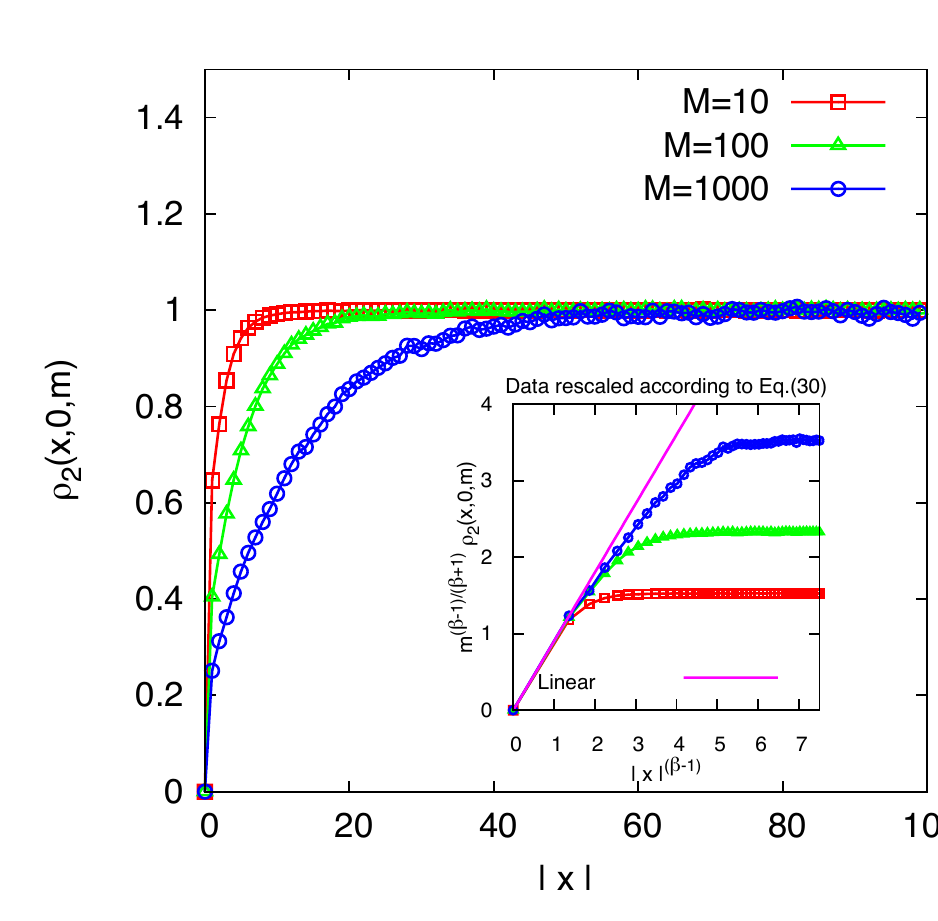}
\caption{Shape of the depletion zone as given by the prediction Eq. (\ref{eq-depletionZoneLevy}) for the L\'{e}vy characteristic $\beta=3/2$. Each plot corresponds to a different choice of mass, $m$, and have been normalised against the corresponding single-point densities. The inset provides plausible evidence for this prediction through the data collapse suggested by Eq. (\ref{eq-depletionZoneLevy}).}

\end{figure}

We may now ask what happens in the case of L\'{e}vy flights. 
Fig.~\ref{fig-anticorrelationVaryBeta} shows that as $\beta$ is varied from
2 to 1, the depletion zone disappears. For comparison, the corresponding
correlation functions for the limiting cases of completely local hops and
completely random hops are shown. The results for  L\'{e}vy flights lie
somewhere in between. Interestingly, the correlation between particles
does not seem to vanish linearly with separation in the  L\'{e}vy case.
See, for example, the main panel of Fig.~\ref{fig-newCorrelationHole}.

In the case of stationary CCA with L\'{e}vy flights, the anomalous diffusion 
length for particles of mass $m$ in one dimension is, dimensionally,  
\begin{equation}
\label{eq-diffusionLengthLevy}
l_{D_\beta} = \left(\frac{D_\beta\,m}{J}\right)^{\frac{1}{1+\beta}}.
\end{equation}
The fact that the correlation between 
particles vanishes linearly in the
case of regular diffusion follows from the calculations performed in
\cite{MRZ2006} and is connected to the fact that the spatial transport mechanism is
diffusion. Dimensionally, any power of 
$\left|\V{x}_i-\V{x}_j \right|/l_D$ in Eq.~(\ref{eq-AA2A}) or 
Eq.~(\ref{eq-rhonCCA}) would 
be consistent so there is no reason to expect that this power should remain
unity in the  L\'{e}vy case when the transport mechanism is changed:
\begin{eqnarray}
\label{eq-rhonCCALevy}
\nonumber
\rho_n(\V{x}_1,\ldots\V{x}_n,m) &\sim& \left(\frac{J}{D}\right)^\frac{n}{1+\beta}\,m^{-\frac{2+\beta}{1+\beta}n}\\
\nonumber
&&\times\prod_{1\leq i < j \leq n} \left[ \frac{\left|\V{x}_i-\V{x}_j\right|}{(\frac{D_\beta\,m}{J})^{1/(1+\beta)}}\right]^\alpha.\\
&&
\end{eqnarray}

We put forward a scaling argument to fix the power of $\alpha$. From the
discussion of the Constant Flux Relation in Sec.~\ref{sec-CFR}, we know that
the second order correlation function at a single spatial point is theoretically
proportional to the mass flux, $J$, in the stationary state regardless of the 
transport mechanism. This fact is strongly supported for the case of L\'{e}vy
flights by the numerical measurements presented in Fig.~\ref{fig-exponents}.
If one accepts that this scaling remains true when the single point
correlation function is split onto two nearby points, then one would expect
that the power of $J$ in Eq.~(\ref{eq-rhonCCALevy}) is unity for $n=2$
(or equally, the power of $m$ should be $-3$) which then requires that the
exponent $\alpha$ should be equal to $\beta-1$. This is consistent with
Eq.~(\ref{eq-rhonCCA}) when $\beta=2$ and consistent with the disappearance
of the depletion zone when $\beta=1$. For $1 < \beta \leq 2$ we
therefore propose the following shape for the depletion zone as measured 
by the two-point function:
\begin{equation}
\label{eq-depletionZoneLevy}
\rho_2(\V{x}_1,\V{x}_2,m) \sim \left(\frac{J}{D}\right)^\frac{2}{1+\beta}\,m^{-\frac{4+2\beta}{1+\beta}} \frac{\left|\V{x}_1-\V{x}_2\right|^{\beta-1}}{(\frac{D_\beta\,m}{J})^{(\beta-1)/(1+\beta)}}.
\end{equation}
This prediction is tested against the numerical measurements for $\beta=3/2$
in Fig.~\ref{fig-newCorrelationHole}. The inset shows the measured values
of $\rho_2(\V{x}_1,\V{x}_2,m)/ \rho_1(m)^2$ for several different values of 
$m$ rescaled by $m^{(\beta-1)/(1+\beta)}$ plotted as a function of
$\left|\V{x}_1-\V{x}_2\right|^{\beta-1}$. According to Eq.~(\ref{eq-depletionZoneLevy}), the data should collapse to a single curve vanishing linearly for
small separations. The collapsed curve seems to rule out the possibility of
linear scaling as one might expect by eye from the main panel of  Fig.~\ref{fig-newCorrelationHole} and makes
the suggested scaling, Eq.~(\ref{eq-depletionZoneLevy}), seem plausible.
Further investigations, preferably supported by analytic calculations,
will be required in order to draw a definitive conclusion. In particular,
it is important to properly characterise the ``internal" mass structure of
correlation functions like Eq.~(\ref{eq-rhonCCA}) for different
masses, particularly for $n=2$. It might be expected that the denominators
of the $\left|x_i-x_j\right|$ terms have mass dependences of the
form $(m_im_j)^{1/6}\,f(m_i/m_j)$ where $f(x)$ is a homogeneous function of
degree 0 which has been ignored in our scaling arguments. The
legitimacy or otherwise of this neglect is likely to be somehow related to the question 
of the locality of the mass cascade (see \cite{CRZ2008} for some discussion).
This is a tricky issue in general and will require further work.

\section{Conclusions}
\label{sec-conclusions}

To summarise, we have studied in detail, the scaling properties of one 
dimensional stationary CCA with L\'{e}vy flights. We have demonstrated that the transition 
from mean field statistics to fluctuation dominated statistics usually 
observed as the physical dimension is tuned through two from above, can be
mimicked in one dimension by varying the characteristic exponent, $\beta$, of 
the jump size distribution of the L\'{e}vy flights. This is physically 
reasonable
since the introduction of long range hops provides a mechanism to weaken
correlations between particles in the system and may erase them entirely
if they are sufficiently frequent. Our predicted values for the scaling exponent
of the mass density as a function of $\beta$ based on dimensional arguments
agreed well with numerical simulations of the underlying stochastic particle
system.

We also provided a direct demonstration of the strong universality of the
scaling exponent of the mass-flux-carrying correlation function in the
stationary state. It is independent of both the physical dimension and the
value of $\beta$ as expected from theoretical considerations.

Finally we performed detailed investigations of the spatial structure of the
depletion zones surrounding heavy particles in stationary CCA with both
regular diffusion and L\'{e}vy flights. Our results indicate that in the
case of regular diffusion, the exact results of \cite{MRZ2006} describing
the multiscaling of correlation functions in the decaying $A+A\to A$ model,
have direct analogues for stationary CCA. In the case of  L\'{e}vy flights, 
our numerical studies suggest that the two-point correlation function
measuring the anti-correlation between particles in the system vanishes 
with a non-trivial fractional power of the separation between the
particles as this separation decreases to zero. Using our knowledge of the
exact scaling for the flux-carrying correlation function we put forward a
scaling argument that this power should be $\beta-1$. This scaling is 
consistent with numerical observations but further efforts will be 
required to make it definitive.

It is not immediately evident how to extend the methods employed in 
\cite{CRZ2006,MRZ2006} to account for L\'{e}vy flights since the fractional
Laplacian entering the propagator of effective field theory seems 
difficult to obtain from the operator representation of the master equation.
A direct calculation from the master equation seems equally daunting. 
Nevertheless we expect that some progress in this direction could and should
be made. This will greatly clarify the scaling arguments presented in this
article.

Several interesting extensions could be made to the model in order to make
it more physically realistic. One would be to introduce mass-dependent
diffusion, for example, and allowing the diffusion rate to decrease as 
a power of the mass, $D(m)\sim m^{-a}$, as one would expect for 
real particles. Another would be to replace the L\'{e}vy flights with
L\'{e}vy walks to account for the fact that physical particles cannot 
instantaneously make long range jumps. We know that neither of these
modifications would affect the scaling exponent, $\gamma_2$ of the two-point 
function which is fixed by the CFR alone. The  value of $\gamma_1$ and 
indeed $\gamma_n$ for $n>2$ are, however, expected to be sensitive to such 
modifications and would considerably complicate the picture.

Another natural extension would be to ask similar questions about the
structure of anticorrelations in the so-called charge model with L\'{e}vy diffusion.
The charge model \cite{TAK1989,MS1993} is a closely related model in which
masses (or charges) can take positive and negative values. When the rates of injection of positive and negative
charges are equal, the flux $J$ is zero on average so that the scaling arguments
presented here fail. Instead of an exact conservation law for
charge one instead has a statistical conservation law for charge squared.
This was used to derive the CFR ($\gamma_2=4$) for the charge model in
\cite{CRZ2007}. Repeating the dimensional arguments leading to Eq.~(\ref{eq-DALevy}) in arbitrary
dimension taking  $J$ to be the average flux of $m^2$ gives
$\gamma_1 = \frac{3 d + \beta}{d+\beta}$
as a first guess for the charge model scaling. In $d=1$ one
has $\gamma_1 = \frac{3 + \beta}{1+\beta}$ which gives the
diffusion limited result $\gamma_1=5/3$ when $\beta=2$ and the
mean-field answer $\gamma_1=2$ when $\beta=1$, both of which are known from
\cite{TAK1989}.  If this dimensional prediction were substantiated by
an additional analytic or numerical study it could be used
to generalise the remaining arguments.

\section*{Acknowledgements}
CC thanks R. Rajesh and O. Zaboronski for many illuminating discussions and 
acknowledges funding from RCUK. JH acknowledges funding from EPSRC.

\section*{Copyright}
\noindent Scaling properties of one-dimensional cluster-cluster aggregation with L\'{e}vy diffusion.

\noindent Journal of Statistical Mechanics: Theory and Experiment \textcopyright\hspace{1mm}Copyright (2010) IOP Publishing Ltd.

\bigskip This is an author-created, un-copyedited version of an article accepted for publication in Journal of Statistical Mechanics: Theory and Experiment. IOP Publishing Ltd is not responsible for any errors or omissions in this version of the manuscript or any version derived from it. The definitive publisher authenticated version is available online at doi: 10.1088/1742-5468/2010/05/.

\end{document}